\documentclass{article}

\usepackage[utf8]{inputenc}
\usepackage{floatrow}
\usepackage{graphicx}
\usepackage[T1]{fontenc}
\usepackage{algorithm}
\usepackage{algorithmic}
\usepackage{url} 
\usepackage{lmodern}
\usepackage{setspace}
\usepackage{listings}
\usepackage{longtable}
\usepackage{amsmath}
\usepackage[english]{babel}
\usepackage[export]{adjustbox}
\RequirePackage{atbegshi}
\ProvideTextCommand{\textasciitilde}{OT1}{\~{}}
\usepackage{color}
\usepackage{todonotes}
\usepackage{subfigure}

\begin{document}

\title{Well-supported phylogenies using largest subsets of core-genes by discrete particle swarm optimization}

\author{Reem Alsrraj, Bassam~AlKindy, Christophe~Guyeux,\\ Laurent~Philippe, and Jean-Fran\c cois Couchot}

\maketitle

\abstract{The number of complete 
chloroplastic genomes increases day after day, 
making it possible to rethink plants phylogeny
at the biomolecular era. Given a set of close 
plants sharing in the order of one hundred of core 
chloroplastic genes, this article focuses on 
how to extract the largest subset of sequences in order to obtain the most supported species 
tree. Due to computational complexity, a discrete
and distributed Particle Swarm Optimization (DPSO) 
is proposed. It is finally applied
to the core genes of \emph{Rosales} order.}

\section{\bf Introduction}

Given a set of biomolecular sequences or characters, various well-established approaches have been developed in recent years to deduce their phylogenetic relationship, encompassing distance-based matrices, Bayesian inference, or maximum likelihood~\cite{Stamatakis21012014}. Robustness aspects of the produced trees can be evaluated too, for instance through bootstrap analyses.
However the relationship between this robustness, the real accuracy of the phylogenetic tree, and the amount of data used for this reconstruction is not yet completely understood. 
More precisely, if we consider a set of species reduced to lists of gene sequences, we have an obvious
dependence between the chosen subset of sequences and the obtained tree (its topology or robustness).
This dependence is usually regarded by the mean
of gene trees merged into a phylogenetic network.

This article investigates the converse approach: it  starts by the union of whole core genes, and tries to remove the ones that blur the phylogenetic signals. More precisely, the objective is to find the largest part of the genomes that produces a phylogenetic tree as supported as possible, reflecting by doing so the relationship of the largest part of the sequences under consideration. 
Due to overwhelming number of combinations to investigate, a brute force approach is a nonsense, which explains why heuristics have been considered. 
The proposal of this research work is thus the application of a Discrete Particle Swarm Optimization (DPSO) that aims at finding the largest subset of core genes producing a phylogenetic tree as supported as possible. A new algorithm has been proposed and applied, in a distributive manner, to investigate the phylogeny of \emph{Rosales} order.

The remainder of this article is constituted as follows. The DPSO
metaheuristic is recalled in the next section.
The way to apply it for resolving problematic
supports in biomolecular based phylogenies
is detailed in Section~\ref{sec:algo}.
The proposed methodology is then applied to
the particular case of \emph{Rosales} in
Section~\ref{sec:exp}.
This article ends by a conclusion section, in 
which the article is summarized and intended future work is outlined.

\section{\bf Discrete Particle Swarm Optimization}
\label{Sec:PSO}
Particle Swarm Optimization (PSO) is a stochastic optimization technique developed by Eberhart and Kennedy in 1995~\cite{kenndy1995particle}. %, and the most recent update of this study is realized on 2010~\cite{kennedy2010particle},
%It is inspired by social behavior of bird flocking or fish schooling, and by social learning. 
PSOs have been successfully applied in function optimization, artificial neural network training, and fuzzy system control. 
In this metaheuristic, particles follow a very simple behavior that is to learn from the success of neighboring individuals. An emergent behavior enables individual swarm members to take benefit from the discoveries or from previous experiences of the other members that have obtained more accurate
solutions. %Optimum particles represent the potential solution in a possibly high dimensional search space. 
%
%PSO is thus a stochastic optimization method that relies on an iterative evolution of a set (the swarm) of candidate solutions in the shape of individuals. Particles move in the solution space and follow the current optimal individual. %
In the case of the standard binary PSO model~\cite{intechopen}, the particle position is a vector of $N$ parameters that can be set as ``yes'' or ``no'', ``true'' or ``false'', ``include'' or ``not include'', \emph{etc.} (binary values).
A function associates to such kind of vector a real number score according to the optimization problem. 
The objective is then to define a way to move the particles in the $N$ dimensional binary search space so that they 
produce the optimal binary vector w.r.t. the scoring function.
%These binary values can also be a representation of a real data in binary search space. In the binary PSO model, best local  and best global solution inside each particle are updated as in the classical model. The major difference between binary PSO and classical model is centered on the velocity expression. 

%A velocity must be generated randomly between 0 and 1. 

In details, each particle $i$ is thus represented by a binary vector $X_i$ (its position). Its length $N$ corresponds to the dimension of the search space, that is, the number of binary parameters to investigate. An $1$ in coordinate $j$ in this vector means that the associated $j$-th parameter is selected.
A swarm of $n$ particles is then a list of $n$ vectors of positions $(X_1, X_2, \dots, X_n)$ together with their associated velocities $V = (V_1, V_2, ..., V_n)$, which are $N$-dimensional vectors of real numbers between 0 and 1. These latter are initially set randomly. %Elements of particle $i$, are arranged from left to right hand side and they correspond to the included real data in a given binary search space. 
At each iteration, 
the new velocity is computed as follows:
\begin{eqnarray}\label{eq:2}
V_i(t+1)= w.V_i(t)+\phi_1(P_{i}^{best}-X_{i})+\phi_2(P_{g}^{best}-X_{i})
\end{eqnarray}
where $w$, $\phi_1$, and $\phi_2$ are weighted parameters setting the level of each 3 trends for the particle, which are respectively to continue in its adventurous direction, to move in the direction of its own best position $P_{i}^{best}$, or to follow the gregarious instinct to the global best known solution $P_{g}^{best}$.
Both $P_{i}^{best}$ and $P_{g}^{best}$ are computed according to the scoring function.

The new position of the particle is then obtained using the equation below:
\begin{eqnarray}\label{eq:3}
X_{ij}(t+1)= 
\begin{cases}
      1, & \text{if}\ {\tiny r}_{ij}\leq Sig(V_{ij}(t+1)) , \\
      0, & \text{otherwise},
    \end{cases} 
\end{eqnarray}
where $r_{ij}$ is a chosen threshold that depends on both the particle $i$ and the parameter $j$, while the $Sig$ function operating as selection criterion is the sigmoid one~\cite{intechopen}, that is:
\begin{eqnarray}\label{eq:1}
Sig(V_{ij}(t))=\frac{1}{1+e^{-V_{ij}(t)}} .
\end{eqnarray}

%\subsection{Discrete Particle Swarm Algorithm Applied to Attribute Selection in Bioinformatics Data Set}

%\begin{color}{red}
%An algorithm deals with discrete variables (attributes). Its population of candidate solutions contains particles of different sizes. The elements of each particle represent the selected attributes. Each attribute is identified by a unique positive integer number. These numbers, vary from 1 to n. These attributes are randomly selected and then they are inserted in a specific particle. Each particle X(i) keeps a record of the best local and global position it has ever attained. The information are sorted in a separated particle labels. Note that the global position is equal to the local position present in the swarm. Each of the n selected elements in the particle has a proportional likelihood which is initialized to 1. Updating velocities depends on three constant factors,are called, $\alpha$, $\beta$ and $\gamma$ parameters chosen by the user used to update the proportional likelihood for the selected elements. Then we multiply these attributes by random binary factors~\cite{GECCO}.
%\end{color}

\section{\bf PSO applied to phylogeny}
\label{sec:algo}
Let us consider, for illustration purpose, a set of chloroplast genomes of \emph{Rosales}, which has already been analyzed in~\cite{genetic2015} using an hybrid genetic algorithm and Lasso test approach. We sampled 9 ingroup species and 1 outgroup (\textit{Mollissima}), see Table~\ref{tab:species} for details, which have been annotated using DOGMA~\cite{RDogma}. We can then compute the core genome (genes present everywhere), whose size is equal to 82 genes, by using for instance the method described in~\cite{Alkindy2014,Alkindy_BIBM2014}. After having aligned them using MUSCLE, we can infer a phylogenetic tree with RAxML~\cite{Stamatakis21012014} (for a general presentation on phylogenetic tree
construction see, \emph{e.g.},~\cite{Review}).
If all bootstrap values are larger than 95, then we can reasonably consider that the \emph{Rosales} phylogeny is resolved, as the largest possible number of genes has led to a very well supported tree.

\begin{table}[ht!]
\caption{\textbf{Genomes information of \textit{Rosales} species under consideration}}
\label{tab:species}
\centering
\begin{tabular}{l l c l l}
\hline\hline
Species&Accession&Seq.length&Family&Genus \\ [0.5ex]
\hline
\textit{Chiloensis}&NC\_019601&155603 bp &\textit{Rosaceae} &\textit{Fragaria}\\
\textit{Bracteata}&NC\_018766&129788 bp &\textit{Rosaceae} &\textit{Fragaria}\\ 
\textit{Vesca}&NC\_015206&155691 bp &\textit{Rosaceae} &\textit{Fragaria}\\
\textit{Virginiana}&NC\_019602&155621 bp &\textit{Rosaceae} &\textit{Fragaria}\\
\textit{Kansuensis}&NC\_023956&157736 bp &\textit{Rosaceae} &\textit{Prunus}\\
\textit{Persica}&NC\_014697&157790 bp &\textit{Rosaceae} &\textit{Prunus}\\
\textit{Pyrifolia} &NC\_015996 &159922 bp &\textit{Rosaceae} &\textit{Pyrus} \\
\textit{Rupicola}&NC\_016921&156612 bp &\textit{Rosaceae} &\textit{Pentactina}\\
%\textit{Soja}&NC\_022868&152217 bp &\textit{Fabaceae} &\textit{Glycine}\\
\textit{Indica}&NC\_008359 &158484 bp &\textit{Moraceae}&\textit{Morus}\\
\textit{Mollissima}&NC\_014674 &160799 bp &\textit{Fagaceae} &\textit{Castanea}\\
\hline
\end{tabular}
\end{table}

In case where some branches are not well supported, we can wonder whether a few genes can be incriminated in this lack of support, for a large variety of reasons encompassing homoplasy, stochastic errors, undetected paralogy, incomplete lineage sorting, horizontal gene transfers, or even hybridization.
If so, we face an optimization problem: \emph{to find the most supported tree using the largest subset of core genes}. Obviously, a brute force approach investigating all possible combinations of genes is intractable ($2^N$ phylogenetic trees for $N$ core genes, with $N=82$ for \emph{Rosales}).

More precisely, genes of the core genome are supposed to be lexicographically ordered.
Each subset $s$ of the core genome is thus associated with a unique binary word $w$ of length $n$: for each $i$, $1\le i \le n$, 
$w_i$ is 1 if the $i$-th core gene is in $s$ and 0 otherwise. 
Any  $n$-length  binary word $w$ can be associated with its percentage $p$ of 1's and the lowest bootstrap $b$
of the phylogenetic tree we obtain when considering the subset of genes associated to $w$. 
Each word $w$ is thus associated with a fitness score value $b+p$.

Let us be back in the PSO context. 
The search space is then $\{0,1\}^N$. 
Each node of this $N$-cube is associated with the set of following data: its subset of core genes, the deduced phylogenetic tree, its lowest bootstrap $b$ and the percentage $p$ of considered core genes, and, finally, the score $b+p$. 
Notice that two close nodes of the $N$-cube have two close percentages of core genes.
We thus have to construct two phylogenies based on close sequences, leading to a high probability to the same topology with close bootstrap. 
In other words, the score remains essentially unchanged when moving from
a node to one of its neighbors.
It allows to find optimal solutions using approaches like PSO. 

%This research treats phylogenetic problems; PSO simulates the behaviors of phylogenetic trees. Suppose the following scenario: a group of genomes are generating random trees. In each iteration, there is only one best tree with a maximum fitness score in the area that is being searched. All the particles do not know which tree is the best. PSO is learned from the scenario and it is used to solve optimization problems. In PSO, each single solution is a ``tree'' in the search space. We call it ``particle''. All the particles have fitness values and velocities which influence the direction of the particles. The objective is to optimize fitness values. The particles direct through the intended space by following the current optimum particle. 

\begin{algorithm}
\caption{\textbf{PSO algorithm}}
\begin{algorithmic} 
\STATE $population \leftarrow 10$, $maxiter \leftarrow 10$
\FOR{each particle in population}
    \STATE $particle[position] \leftarrow [randint(0,1) \text{ for each gene in core genome}]$
   \STATE $particle[velocity] \leftarrow [rand(0,1) \text{ for each gene in core genome}]$
    \STATE $particle[score] \leftarrow 0$
    \STATE $particle[best] \leftarrow \text{Empty list}$
\ENDFOR 
\WHILE{$fitness < b+p$ and $iter< Maxiter$}
\FOR{each particle in population}
\STATE Calculate $new\_fitness$
\IF {$new\_fitness > fitness$}
\STATE $particle[score] \leftarrow new\_fitness$
\STATE $particle[best] \leftarrow particle[position]$
\ENDIF
\ENDFOR
\STATE $fitness \leftarrow  max(particle[score])$
\STATE $Gbest \leftarrow position[Max (Particle[score]\text{in population)}]$
\FOR{each particle in population}
\STATE Calculate particle velocity according to Equation~\eqref{eq:2}
\STATE Update particle position according to Equations~\eqref{eq:1} and~\eqref{eq:3}
\ENDFOR
\ENDWHILE
\end{algorithmic}
\label{algo:PSO}
\end{algorithm}

Initially, the $L$ (set to 10 in our experiments) particles are  
randomly distributed among all the vertices (binary words) of the $N$-cube 
that have a large percentage of 1.
The objective is then to move these particles in the cube, hoping that they will converge to an optimal node. At each iteration, the particle velocity is updated according to the fitness and its best position.
It is influenced by constant weight factors according to Equation~\eqref{eq:2}.
In this one, we have set $c_1 = 1$, $c_2 =1$, while $r_1$, $r_2$ are random numbers between (0.1,0.5), and $w$ is the inertia weight. 
This latter determines the contribution rate of a particle's previous velocity to its velocity at the current time step. To increase the number of included components in a particle, we reduced the interval of Equation~(2) to [0.1, 0.5]. For instance, if the velocity $Vi$ of an element is equal to 0.511545 and $r=0.83$, then $Sig(0.51)=0.62$. So $r>Sig(Vi)$ and this will lead to 0 in the vector elements of the particle. By minimizing the interval we increase the probability of having $r<Sig(Vi)$, and this will lead to more 1s, which means more included elements in the particle.
%Researchers in many articles~\cite{IEEE,SwarmIntell} %, and ~\cite{INJ} stated that
A large inertia weight facilitates a global search while a small inertia weight tends more to a local investigation~\cite{SwarmIntell}. %Further, a similar change is made from the PSO. Presenting how much the amount of memory from the previous position will affect the new velocity. 
%Indeed the velocity increases with time when $w > 1$, thus the particles will accelerate to maximum velocity, and so the swarm will be divergent. Conversely, if $w < 1$, then the particle velocities will decrease until reaching zero. To say this differently, 
A larger value of $w$  facilitates a complete exploration, whereas small values promote exploitation of areas. 
This is why Eberhart and Shi suggested to decrease $w$ over time, typically from 0.9 to 0.4,  thereby gradually changing from exploration to exploitation.
Finally, each particle position is updated according to % calculates velocity. Equation(~\ref{eq:1}) and 
Equation~\eqref{eq:3}, see Algorithm~\ref{algo:PSO} for further details. In this algorithm, the particle is defined by its position (a binary word) in the cube together with its velocity (a real vector). %) update particle position.%~\cite{Swarmwebsite}. %Algorithm \ref{algo:init} illustrates the initialization of population.

\section{\bf Experimental results and discussions}
\label{sec:exp}
\begin{figure}
\CenterFloatBoxes
\begin{floatrow}
\ffigbox
  {\includegraphics[scale=0.4]{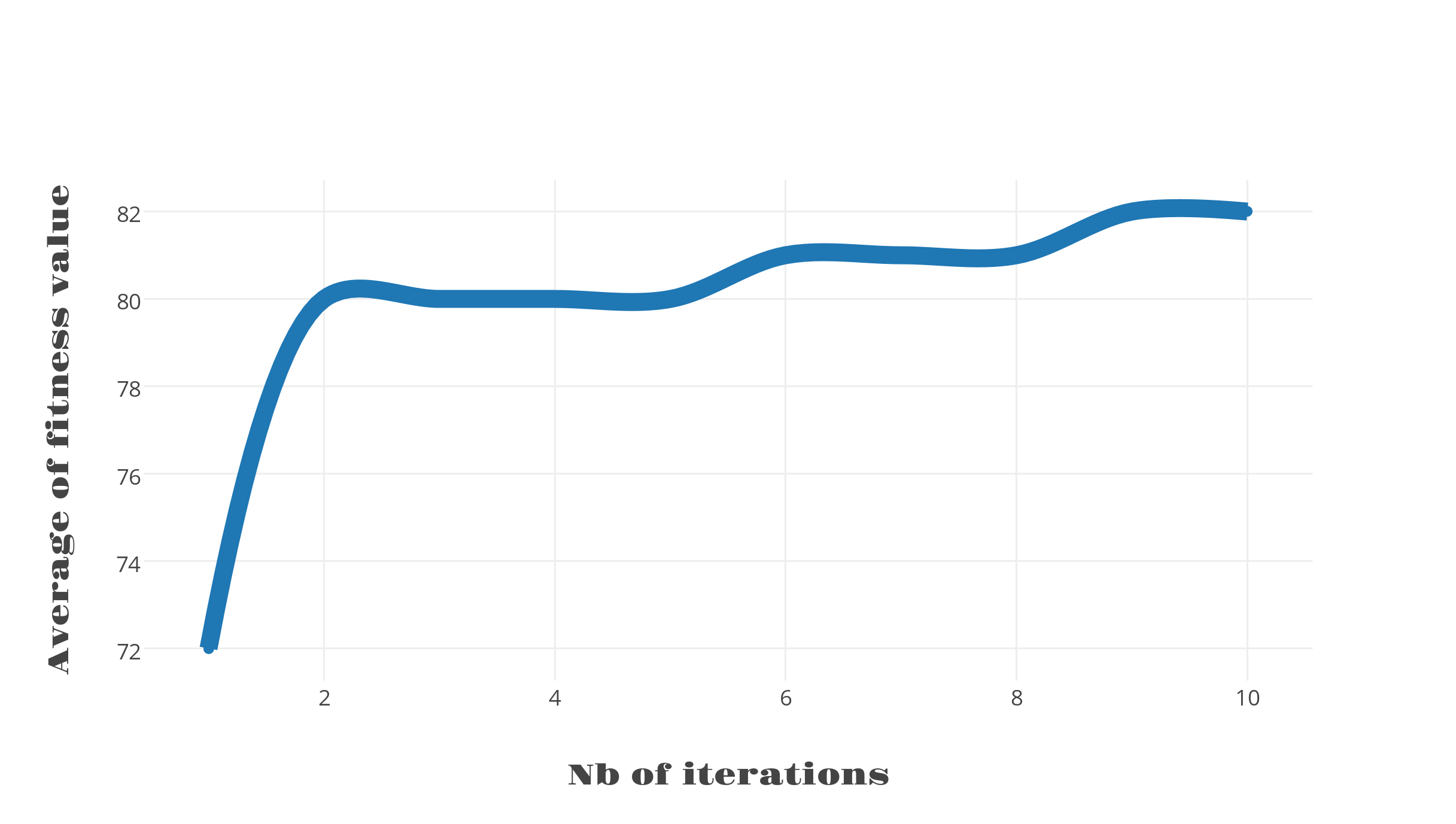}}
  {\caption{\textbf{Average fitness of \textit{Rosales} order}}\label{overflow first version}}
\killfloatstyle
\ttabbox
  {\footnotesize
  \begin{tabular}{c | c| c | c  } 
&Removed & &  \\
Swarm & genes & $(p+b)/2$ & $b$ \\ [0.5ex]
\hline %\hline
1 &4 &75.5&73\\
2 &6  &75.5&76\\
3 &20 &75  &88 \\
4 &52 &59.5&89 \\
5 &3  &75.5&72 \\
6 &19 &77.5&92 \\
7 &47 &63.5&92 \\
8 &9 &73.5&74 \\
9 &10 &72.5 &73 \\
10&13 &76.5 &84 \\
\end{tabular}
  }
{\caption{Best tree in each swarm\label{tab:table2}}}
\end{floatrow}
\end{figure}

We have implemented the proposed DPSO algorithm on the \emph{M\'esocentre de calculs} supercomputer facilities of the University of Franche-Comt\'e. Investigated \emph{Rosales}
species are listed in Table~\ref{tab:species}. % which is a set of computers interconnected by a high-speed network to obtain computing capabilities of several TFLOPS, accompanied by a large storage capacity. Based on Linux Redhat distribution adapted by Bull Xbas.
10 swarms having a variable number of particles have been launched 10 times, with $c_1=1, c_2=1$, and $w$ 
 linearly decreasing from 0.9 to 0.4. %Our algorithm has been tested and compared on the following data sets: \textit{Rosales, Malpighiales}, and \textit{Caryophyllales}. Table~\ref{tab:table1} shows the parameters and values that have been used.
%\begin{table}[ht]
%\centering
%\caption{Binary PSO parameter settings}\label{tab:table1}
%\begin{tabular}{c c c c} 
%\hline\hline 
%population &c1 &c2 &w \\
%\hline
%10 &1 & 1 & 0.9-0.4\\
%2&10 & 2.05 & 2.05 &-&0.0-0.729 \\
%\hline 
%\end{tabular}
%\end{table}
%
%In this experiment, $w_{max}$ is set to 0.9 and $w_{min}$ is equal to 0.4, in such a way that
%inertia weight $w$ ranges between 0.4 and 0.9. 
%More precisely, at
%iteration number $iter$, $w$ is computed using Equation~\eqref{eq:5} from~\cite{premalatha2009hybrid}:
%\begin{eqnarray}\label{eq:5}
%w= w_{max}-\frac{(w_{max}-w_{min})}{maxIter} \times iter
%\end{eqnarray}
%in which $maxIter$ is the maximum number of iterations that has been initially chosen. 
%\JFC{Linearely decreasing <=> w= w_{max}-\frac{(w_{max}-w_{min})}{maxIter} \times iter}
%The inertia weight decreases then gradually after each iteration until convergence. 
%For both versions, we have obtained at least 60 trees at each execution. 
%Information of the best obtained topologies are then extracted from 
%the number of computed trees and on the best minimal bootstrap values for each tree. %We apply a specific algorithm to view the best tree with the minimum bootstrap score.\\
%
%
%We launched 10 experiments on our supercomputer facilities and with the \textit{Rosales} order of Table~\ref{tab:species}. Each experiment were constituted by 10 swarms, each swarm having 10 particles. 
Obtained results are summarized in Table~\ref{tab:table2} that contains, for each 10 runs of each 10 swarms: the number of removed genes and the minimum bootstrap of the best tree. Remark that some bootstraps are not so far from the intended ones (larger than 95), whereas the number of removed genes are in average larger than what we desired.

\begin{figure}[!hb]
\centering
\subfigure[$Topology_0$]{\includegraphics[width=70mm]{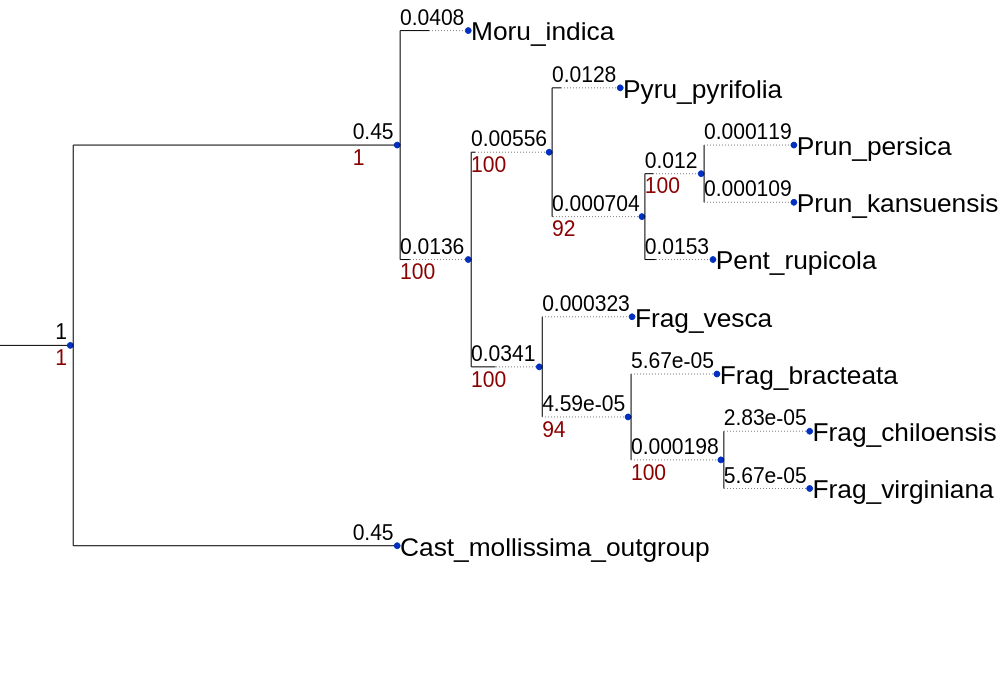}\label{Topology0}}
\subfigure[$Topology_4$]{\includegraphics[width=70mm]{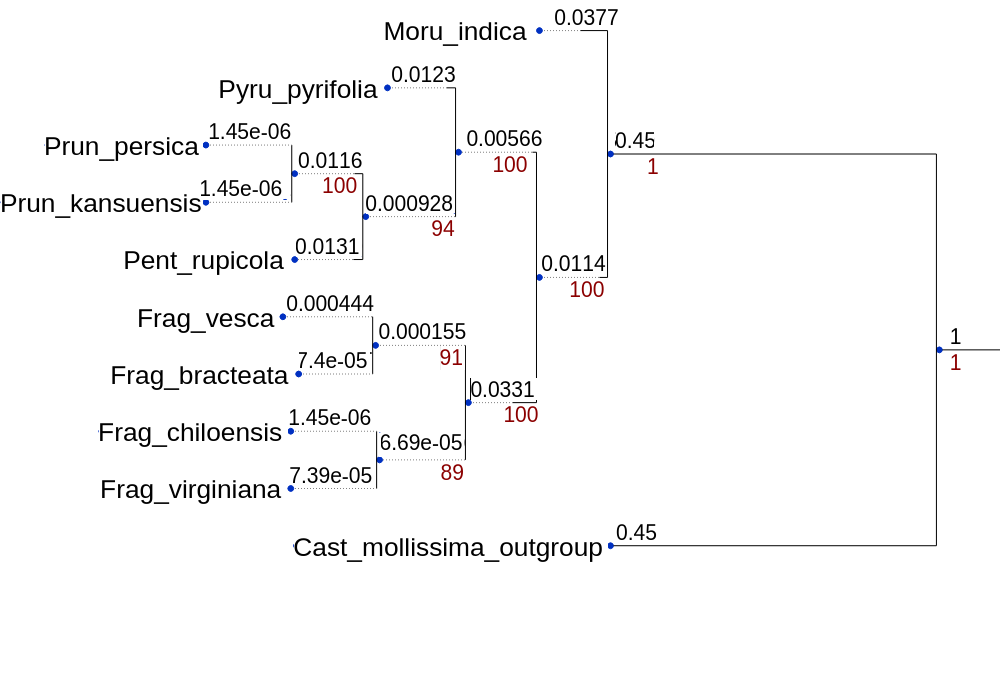}\label{Topology4}}
\subfigure[$Topology_2$]{\includegraphics[width=65mm]{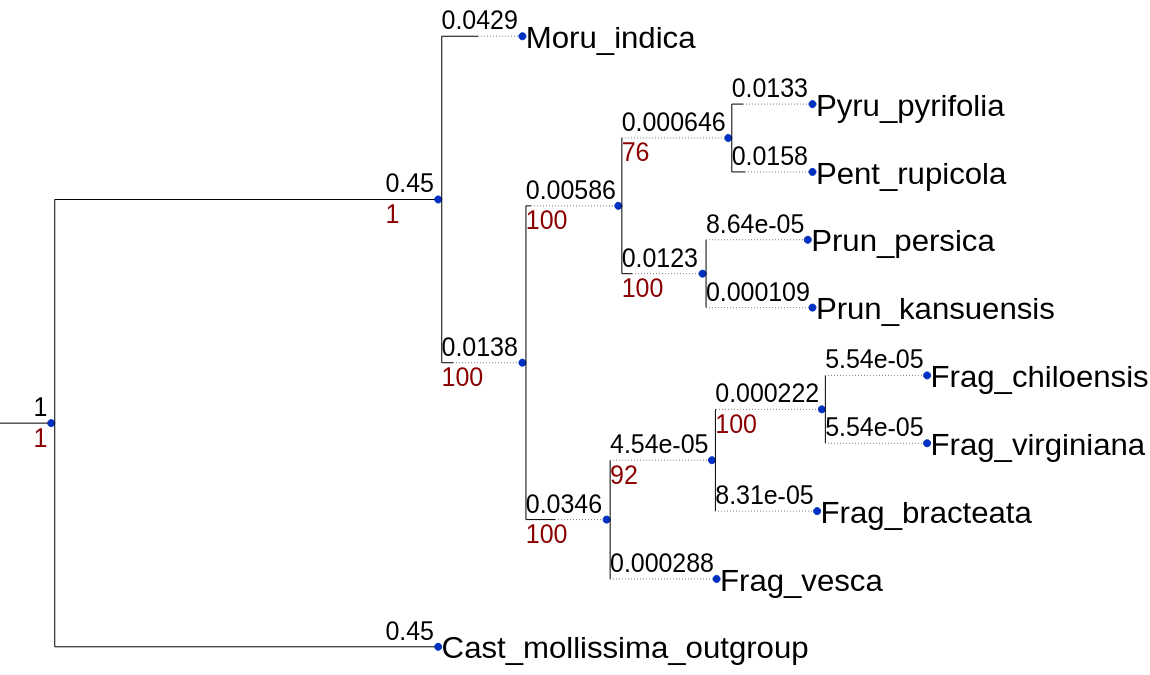}\label{Topology2 }}
\caption{\textbf{The best obtained topologies for \emph{Rosales} order}}
\label{fig:topo}
\end{figure}

7 topologies have been obtained after either convergence or $maxIter$ iterations. Only 3 of them have occurred a representative number of time, namely the Topologies 0, 2, and 4, which are depicted in Figure~\ref{fig:topo} (see details in
Table~\ref{tab:table3}). 
These three topologies are almost well supported, except in a few branches. We can notice that the differences in these topologies are based on the sister relationship of two species named \emph{Fragaria vesca} and \emph{Fragaria bracteata}, and of the relation between \emph{Pentactina rupicola} and \emph{Pyrus pyrifolia}. %In $topology_0$, they are ranked in different levels, while they are ranked in the same level in $topology_4$. 
Due to its larger score and number of occurrences, we tend to select Topology~0 as the best representative of the \emph{Rosale} phylogeny. %From Table~\ref{tab:table3}, we can select $topology_0$ as the one who meets its biological meaning, because it has the maximum number of occurrences and bootstrap values with low number of removal genes. But of course, we still need a tool to support this hypothesis. %\color{black}  % below shows these topologies and their information. The topologies are showed in Figures \ref{Topology0 first method}, \ref{Topology2 first method} and \ref{Topology3 first method}.

\begin{table}[!ht]
\caption{\textbf{Best topologies obtained from the generated trees.} $b$ is the lowest bootstrap of the best tree having this topology, while $p$ is the number of considered genes to obtain this tree.}\label{tab:table3}
\centering
\scalebox{0.8}{
\begin{tabular}{c|c | c | c | c } 
%\hline\hline
%& &&&& &Removed\\[0.5ex]
Topology & Swarms & $b$ & $p$ &Occurrences\\[0.5ex]
\hline %\hline
0 & 1, 2, 3, 4, 5, 6, 7, 8, 9, 10      &92 & 63 &568 \\
1 & 1, 2, 3, 4, 5, 6, 10              &63  &45  &11  \\
2 & 1, 2, 3, 4, 5, 6, 7, 8, 9, 10     &76  &67 &55  \\  
3 & 8, 1, 2, 3, 4                        &56  &41  &5   \\
4 & 1, 2, 3, 4, 5, 6, 7, 8, 9, 10     &89 & 30 &65  \\ 
5 & 1, 3, 4, 5, 6, 9                     &71  &33  &9   \\
6 & 5, 6                                 &25  &45  &2   \\
%\hline\hline
\end{tabular}}
% \end{minipage}
\end{table}

% \begin{figure}
% \centering
% {\includegraphics[width=90mm]{v3}}
% \caption{Average fitness of \textit{Rosales} family\label{overflow first version}}
% \end{figure}

To further validate this choice, %We then used 
consel~\cite{shimodaira2001consel} software has been used on per site likelihoods of each best tree obtained using RAxML~\cite{Stamatakis21012014}. Consel % in order to give a confidence tree from given phylogenetic trees by 
ranks the trees after having computed %depending on computing 
the $p$-values of various well-known statistical tests, like the so-called approximately unbiased (au), Kishino-Hasegawa (kh), Shimodaira-Hasegawa (sh), and Weighted Shimodaira-Hasegawa (wsh) tests. %best tree computations. 
%To do so, the 3 best trees have been sent to  RAxML~\cite{Stamatakis21012014}, together with the alignment file, the GTRGAMMA mutation model, and the \emph{-f G} option. We obtained, by doing so, the per site likelihood according to each topology, and this latter has been sent to consel. 
Obtained results are provided in Table~\ref{tab:consel1}, they confirm the selection of Topology~0 as the tree reflecting the best the \emph{Rosales} phylogeny.

\begin{table}[!ht]
\centering
\caption{\textbf{Consel results regarding best trees}\label{tab:consel1}}
\scalebox{0.9}{
\begin{tabular}{c|c|c|c|c||c|c|c|c|c|c}
%\hline %\hline
Rank & item &   obs &     au &     np &     bp &     pp &     kh &     sh &    wkh &    wsh \\ \hline
1 & 0 & -1.4 & 0.774 & 0.436 & 0.433 & 0.768 & 0.728 & 0.89 & 0.672 & 0.907 \\ 
2 & 4 & 1.4 & 0.267 & 0.255 & 0.249 & 0.194 & 0.272 & 0.525 & 0.272 & 0.439 \\ 
3 & 2 & 3 & 0.364 & 0.312 & 0.317 & 0.037 & 0.328 & 0.389 & 0.328 & 0.383 \\ 
\end{tabular}}
\end{table}

%\color{red}
%From Table~\ref{tab:consel1}, several well known statistical tests such as (\emph{bootstrap probability} (BP), \emph{Shimodaira-Hasegawa} (SH), and \emph{Weighted Shimodaira-Hasegawa} (WSH)) are used by consel to give a confidence measure for a set of candidate trees in order to select a confidence one. The procedure is simple which starts by computing the \emph{p}-value from maximum likelihood (ML) model (\emph{i.e.}, GTR model of RAxML) based on different bootstrap replications, then candidate trees are ranked based on computed minimal ML values. For each given tree, statistical methods are then used to compute the probability value (between 0 and 1) from bootstrap replications and select the tree with greater \emph{p}-values. We can notice this latter in the tree provided by $topology_0$ which has higher \emph{p}-values than $topology_1$. 
%\color{black}
%We get 778 trees(after Deleting frequencies). We obtained 7 topologies which are unique. But, we kept only these which have good minimum bootstrap and less omitted genes.

\section{\bf Conclusion} % and future work}

A discrete particle swarm optimization algorithm has been proposed in this article, which focuses on the problem to extract the largest subset of core sequences with a view to obtain the most supported phylogenetic  tree. This heuristic approach has then been applied to the 82 core genes of the \emph{Rosales} order.

%In future work, the authors' intention is to investigate a large set of PSO variants, by systematically testing the effects of all parameters, and by modifying the velocity equation. We will deeply compare various meta-heuristics approaches like simulated annealing, to determine the ones that answer the best to our problem under consideration. Finally, these algorithms will be applied on a large set of plant species, while a nucleus version will be investigated.

\bibliographystyle{unsrt}
\bibliography{biblio}
\end{document}